\documentclass[fleqn,twoside]{article}
\usepackage{espcrc2}

\newcommand{\be}{\begin{equation}}
\newcommand{\ee}{\end{equation}}
\newcommand{\bea}{\begin{eqnarray}}
\newcommand{\eea}{\end{eqnarray}}

\usepackage{graphicx}
\usepackage[figuresright]{rotating}

\title{Light quark masses from Domain Wall Fermions}

\author{C. Dawson
\address{RIKEN-BNL Research Center,Bldg 510a, Upton, NY 11973-5000}
[RBC Collaboration]
\thanks{We thank RIKEN, Brookhaven National Laboratory and the U.S.\ Department
of Energy for providing the facilities essential for the completion of
this work.}}

\begin{document}

\begin{abstract}
We present results for the renormalised light and strange quark masses
calculated using Domain Wall Fermions in quenched QCD. New results using the
DBW2 gauge action at $a^{-1} \approx 2 {\rm GeV}$ and $a^{-1} \approx 1.3 {\rm
GeV}$ will be presented and compared against existing results at $a^{-1}
\approx 2GeV$ using the Wilson gauge action. This comparison allows a study of
the uncertainties due to both finite lattice spacing and residual chiral
symmetry breaking effects.
\vspace{1pc}
\end{abstract}


\maketitle

\section{INTRODUCTION} 

We report here results for the light and strange quark masses extracted from
the quark mass dependence of the meson spectrum calculated using domain wall
fermions \cite{Kaplan,Furman:1995ky} (DWF), see \cite{Blum:2000kn} for our notation
and conventions, in the quenched approximation and renormalised using the
non-perturbative renormalisation (NPR) technique developed by the
Rome-Southampton group \cite{Martinelli:1995ty}.

All the results presented will be from simulation using a lattice with
spatial dimensions $16^3\times32$ and a fifth dimension of size 16.  The new
results use the DBW2 gauge action \cite{deForcrand:1999bi} at $\beta=0.87$ and
$\beta=1.04$ corresponding to $a^{-1}\approx 1.3{\rm GeV}$ and $a^{-1} \approx
2 {\rm GeV}$ respectively. This action is chosen for its much smaller explicit
chiral symmetry breaking when compared to other actions\cite{Orginos:2001xa}.
The results will be compared  with those of previous studies using the Wilson
gauge action with $\beta=6.0$ \cite{Blum:2000kn,Blum:1999xi,Blum:2001sr}. For
$\beta=0.87$, $M_5=1.8$ was used and 53 and 100 configurations were collected
for the NPR and hadron spectrum respectively, whereas for $\beta=1.04$, 
$M_5=1.7$ was used and 51 and 405 configurations were collected.


\section{RENORMALISATION}

Our final result will be the renormalised mass in the $\overline{MS}$-scheme
at 2 GeV. The relation between this value and the bare mass, $m_f$, consists
of several factors and may be written 
\be m_{\rm ren} = Z^{\rm match} Z_m^{RI}
\left( m_f + m_{\rm res}\right) \, .  
\ee 
The residual mass, $m_{\rm res}$, occurs because of small explicit chiral
symmetry breaking effects due to the finite extent of the fifth dimension.
This we determined from the breaking term of the axial Ward-Takahashi identity
\cite{Blum:2000kn}.  $Z_m^{RI}$ is the mass renormalisation factor in the
RI/MOM-scheme which is calculated directly on the lattice using the method of
the Rome-Southampton group \cite{Martinelli:1995ty} and then matched to the
$\overline{MS}$-scheme at 2 GeV using a two-loop perturbative matching factor,
$Z^{\rm match}$ \cite{Chetyrkin:1999pq}.

For the NPR technique to be successful a ``window'' of momenta must
exist for which contamination due to (low-momenta) condensate effects are
small but also for which (high-momenta) lattice artifacts are
suppressed. Previous work shows that when using DWF with the Wilson
action at $a^{-1} \approx 2 {\rm GeV}$ this window exists\cite{Blum:2001sr}. In
this work we confirmed that this is also the case for the DBW2 action and
checked the feasibility of NPR at $a^{-1} \approx 1.3{\rm GeV}$.

While $Z_m$ may be calculated directly from the momentum-space quark
propagator this approach has large systematic errors.  Instead we consider the
renormalisation of the flavour non-singlet local quark bilinear operator,
$\overline{u}\Gamma d$. The NPR technique provides a clean extraction of the
ratio $Z_\Gamma/Z_q$, where $Z_q$ is the quark renormalisation factor. Putting
the value of $Z_S/Z_q$ (scalar) and $Z_A/Z_q$ (axial) together with the value
of $Z_A$ calculated from the hadronic matrix element of the local and
conserved axial currents with the pseudo-scalar density\cite{Blum:2000kn}
gives our final value for $Z_S$. This can then be used to calculate
$Z_m$ as $Z_S = 1/Z_m$.

\begin{figure}[!t]
\begin{center}
\resizebox{7.2cm}{!}{
\rotatebox{-90}{
\includegraphics{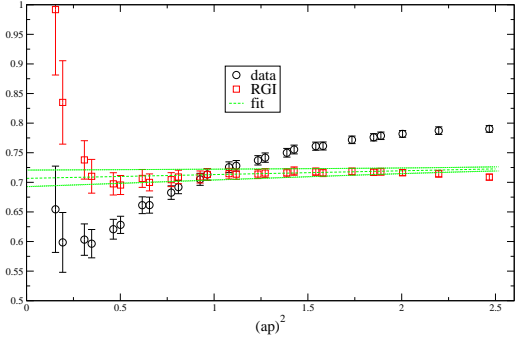}}}
\end{center}
\vspace{-1cm}
\caption{$Z_S/Z_q$ bare and RGI for the DBW2 action at $\beta=1.04$}
\vspace{-0.5cm}
\label{dbw2_2_rgi}
\end{figure}

An important check of the existence of a ``window'' may be made by dividing
$Z_S/Z_q$ by the predicted running from 3-loop perturbation theory to produce
a renormalisation group invariant (RGI) value. Fig.~\ref{dbw2_2_rgi} shows
this for $\beta=1.04$ DBW2 where the RGI and bare data are constrained to
agree at $(ap)^2=1$. 

\begin{figure}[!t]
\begin{center}
\resizebox{7.2cm}{!}{
\rotatebox{-90}{
\includegraphics{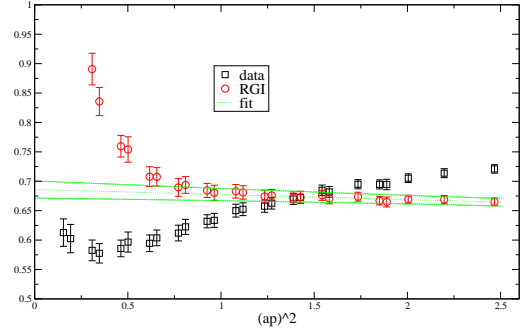}}}
\end{center}
\vspace{-1cm}
\caption{$Z_S/Z_q$ bare and RGI for the DBW2 action at $\beta=0.87$}
\vspace{-0.5cm}
\label{dbw2_1.3_rgi}
\end{figure}

Fig.~\ref{dbw2_1.3_rgi} show this for $\beta=0.87$ DBW2 (in this case the
bare and RGI are constrained to agree at $(ap)^2=1.5$).  There is both more
evidence of curvature at low-momenta due to condensate effects and a noticeable
slope versus $(ap)^2$ at high momenta. Interpreting this slope as $O(a^2)$
contamination and fitting the RGI value to $A + B(ap)^2$ allows the final
value for $Z_S/Z_q$ to be extracted. 

Table \ref{npr_res} collects together the (preliminary) DBW2 results for $Z_A$
and $Z_S/Z_A$ (the latter in the $\overline{MS}$-scheme at 2 GeV), and the
results for Wilson gauge results already quoted in \cite{Blum:2001sr}.


\begin{table}[!b]
\caption{Results for $Z_A$ and $Z_S/Z_A$}
\begin{tabular}{cccc} \hline
Action  & $\beta$    & $Z_A$     & $Z_S/Z_A$     \\ \hline
DBW2    & 0.87 & 0.7776(5) & 0.898(16)(90) \\
DBW2    & 1.04 & 0.8402(2) & 0.849(15)(40) \\
Wilson  & 6.00 & 0.7555(3) & 0.830(09)(35) \\ \hline
\end{tabular}
\label{npr_res}
\end{table}

\section{MESON MASSES}

For the pseudo-scalar meson masses chiral perturbation theory to first
order tells us that
\bea
m_\pi^2 a^2 &=& B_\pi a \overline{m}\label{eq1} \\
m_K^2 a^2   &=& B_\pi a(m_s + \overline{m})/2 \label{eq2}
\eea
where $\overline{m} = \frac{1}{2}\left(  m_u + m_d \right)$, while
for the vector mesons the leading dependence should
be linear in the quark masses
\bea
m_\rho a    &=& A_\rho + B_\rho \overline{m} \label{eq3} \, .
\eea
Our approach to calculating the quark masses is to first fit the 
meson spectrum for the coefficients $B_\pi$, $A_\rho$ and
$B_\rho$ and then to use the physical values for $m_\pi$, $m_K$ and
$m_\rho$ to calculate $a$, $\overline{m}$ and $m_s$.

In extracting the meson masses we must deal with two systematic effects due to
working in the quenched approximation: the contamination of meson correlators
due to the presence of un-suppressed topological near zero-modes
\cite{Blum:2000kn}, and significant deviations from Eq.~\ref{eq1} at light
masses due to quenching effects.

The effects of zero-modes are most apparent at light masses and small volumes,
but also are strongly dependent on which correlators are used to calculate the
mass. In particular when extracting the pseudo-scalar meson mass the zero-mode
effects may be reduced (but not completely avoided) by calculating the mass
from axial correlators rather than the more usual pseudo-scalar
correlators.  In addition to using the axial correlator for the
pseudo-scalar mass extraction, we only use bare masses greater than
$m_f=0.01$ in our analysis.

\begin{figure}[!t]
\begin{center}
\resizebox{7.2cm}{!}{
\rotatebox{-90}{\includegraphics{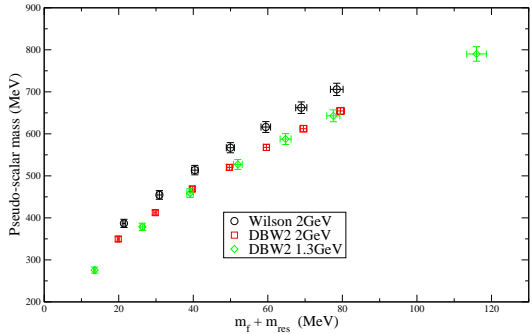}}}
\end{center}
\vspace{-1cm}
\caption{Pseudo-scalar mass versus $m_f+m_{\rm res}$}
\vspace{-0.5cm}
\label{picmp_un_de}
\end{figure}

\begin{figure}[!t]
\begin{center}
\resizebox{7.2cm}{!}{
\rotatebox{-90}{\includegraphics{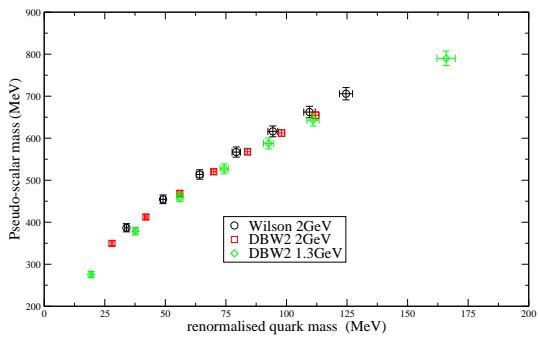}}}
\end{center}
\vspace{-1cm}
\caption{Pseudo-scalar mass versus renormalised mass}
\vspace{-0.5cm}
\label{picmp_de}
\end{figure}

Using the mass extracted from the axial correlator, the consistency of our
different data-sets with each other and the NPR calculation may be
tested. Fig.~\ref{picmp_un_de} shows the pseudo-scalar mass versus $\tilde{m}=
m_f + m_{\rm res}$, both measured in MeV, whereas Fig.~\ref{picmp_de} shows
the same data versus renormalised mass. As can be seen once the
renormalisation factors are taken into account the data-sets show excellent
consistency both between the two gauge action at the same lattice spacing and
the two different lattice spacing with the same gauge action.
The deviation from Eq.~\ref{eq1} is very apparent in our data with a
simple linear fit of
\be (am_\pi)^2 = A_\pi + B_\pi a \tilde{m} 
\ee
for the $\beta=1.04$ DBW2 data having a constant term, $A_\pi$, inconsistent
with zero by 3 standard deviation whereas the intercept of a more general
quadratic fit misses zero by 7 standard deviations. 

As this is the case we include a quenched chiral logarithm in our fitting
formula (but drop its effects for the final calculation of the quark masses).
\be \left( a m_\pi \right)^2 = B_\pi \tilde{m} \left( 1 - \delta \log \left(
\frac{B_\pi \tilde{m} }{\Lambda^2} \right) \right) 
\ee 
where we use a chiral scale of $\Lambda = 1 {\rm GeV}$ \cite{Blum:2001xb}.
This gives very consistent values of $\delta=5.1(19)\times10^{-2}$ and
$\delta=4.6(18)\times 10^{-2}$ for Wilson and DBW2 at 2GeV and
$\delta=2.28(12)\times 10^{-2}$ at 1.3GeV. However, none of the fits are of
high quality and the best treatment of the quenching effects is an issue that
requires further study.

\section{RESULTS}

Putting together the NPR results and the fits to the meson spectrum gives
values for the quark masses in $\overline{MS}$-scheme at 2 GeV. Using
Equations \ref{eq1} to \ref{eq3} implies that 
\be
m_s/\overline{m} = 2 m_K^2/m_\pi^2 - 1 \approx 26
\ee
exactly for our results so we only quote 
the strange quark mass. For this we get $132.5(31)$ MeV ($\beta=0.87$,
DBW2), $132.6(21)$ MeV ($\beta=1.04$, DBW2), and $m_s = 125.6(28)$
MeV ($\beta=6.0$, Wilson) where the errors quoted are purely statistical.

\end{document}